\documentclass[useAMS,usegraphicx]{mn2e}
\title[Dynamical evolution of hot spots in radio loud AGNs]
{On the dynamical evolution of hot spots 
in powerful radio loud AGNs}
\author[N. Kawakatu and M. Kino]{N. Kawakatu$^{1,2}$
\thanks{E-mail:kawakatu@th.nao.ac.jp} and M. Kino $^{1,3}$
\thanks{E-mail:kino@vega.ess.sci.osaka-u.ac.jp}\\
$^{1}$ SISSA, via Beirut 2-4, 34014, Trieste, Italy\\
$^{2}$ National Observatory of Japan, 181-8588 Mitaka, Japan\\
$^{3}$ Department of Earth and Space Science, Osaka University, 
560-0043 Toyonaka, Japan}
\begin{document}

\date{}

\pagerange{\pageref{firstpage}--\pageref{lastpage}} \pubyear{2006}

\maketitle

\label{firstpage}

\begin{abstract}
We describe the dynamical evolution of hot spots velocity, 
pressure and mass density in radio loud active galactic nuclei (AGNs), 
taking proper account of 
(1) the conservations of the mass, momentum, and kinetic energy 
flux of the unshocked jet, 
(2) the deceleration process of the jet by shocks, and 
(3) the cocoon expansion without assuming the 
constant aspect ratio of the cocoon.  
By the detailed comparison with two dimensional relativistic 
hydrodynamic simulations, 
we show that our model well reproduces the whole evolution of 
relativistic jets. 
Our model can explain also the observational trends of the velocity, 
the pressure, the size, and mass density of hot spots 
in compact symmetric objects (CSOs) and FR II radio galaxies. 

\end{abstract}

\section{Introduction}

Which evolutionary tracks are radio loud AGNs (radio galaxies) passing through?
This is one of the primal issues in the study of AGNs (Ryle \& Longair 1967; Carvalho 1985; Fanti et al. 1995; De Young 1997). 
%
%
Stimulated by the observational progress (e.g., Turland 1975; Readhead, Cohen \& Blandford 1978; Bridle \& Perley 1984), a number of hydrodynamic simulations of jet propagations have been performed to examine their physical state of the jet (e.g., Norman et al. 1982; Wilson \& Scheuer 1983; Smith et al. 1985; Clarke, Norman \& Burns 1986; Lind et al. 1989; Clarke, Harris \& Carilli 1997; Marti et al. 1997). 
These numerical studies have confirmed that the jet is composed of ``light'' (i.e., lower mass density) materials compared with an ambient medium to reproduce the observed morphology of the expanding {\it cocoon} (e.g., Norman et al. 1982). However it is hard to examine the whole duration of powerful radio loud AGNs with sufficiently large dynamical range because of the limitation of computational powers.

A new population of 
radio sources  so-called ``compact symmetric objects (CSOs)''
has been recently noticed. The CSOs was first identified by 
Philips $\&$ Mutel (1980, 1982) and  
more complete sample were presented 
by Wilkinson et al. (1994) and Readhead et al. (1996a, b). 
%
%
Concerning the origin of CSOs, two scenarios were initially proposed. 
One is so-called ``frustrated jet scenario''
in which the ambient medium is so dense 
that jet cannot break its way through, so sources are old and confined 
(van Breugel, Miley \& Heckman 1984). 
The other is  ``youth radio source scenario''
in which CSOs are the young 
progenitor of FR II radio galaxies (e.g., Shklovsky 1965; Philips \& Mutel 1982; Carvalho 1985; Fanti et al. 1995; Begelman 1996; 
Readhead et al. 1996a; O'Dea \& Baum 1997).
 Recent observations reveal that their speeds are better understood within 
youth radio source scenario because of their age with $10^{3-5}$yr, 
 which is much  shorter than the age of  FR II sources with $10^{6-7}$yr 
 (e.g., Owaiank, Conway \& Polatidis 1999; Murgia et al. 1999; Taylor et al. 2000). This indicates the possibility of CSOs as the progenitor of FR II sources although their evolutionary tracks are poorly understood.

The hot spot, which is identified as the 
reverse shocked region of the decelerating jet,
is one of the most important ingredients in the 
whole jet system.
The evolution of the hot spot is tightly linked to 
that of cocoon because the cocoon is 
consist of the shocked plasma escaped
from the hot spot (see Fig. 1). 
Observationally, the correlations between the hot spot properties (the velocity, the pressure, the size and the mass density) and projected linear size have been reported for CSOs and FR II sources (Readhead et al. 1996a; Jeyakumar \& Saikia 2000; Perucho \&  Mart\'{i} 2002 ). These observational trends would also reflect the evolutionary tracks of radio loud AGNs. Thus, in order to clarify the physical relation between CSOs and FR II sources, it is inevitable to model the dynamical evolution of hot spots in radio loud AGNs. However, little attention has been paid to this point in spite of lots of theoretical evolutionary models have been proposed based on cocoon dynamics (e.g., Falle 1991; Begelman 1996; Kaiser \& Alexander 1997). 
Thus, the goal of this paper is to construct 
an appropriate dynamical model of hot spots in the radio loud AGNs. 

The plan of the paper is organized as follows. 
In $\S 2$ we outline and model a dynamical evolution of hot spots connected with the cocoon dynamics. In $\S 3$, we compare with previous theoretical and observational works. Conclusions are given in $\S 4$.

\section{Dynamical evolution of hot spots connected with the cocoon expansion}

\subsection{Outline}

In this paper, we model a dynamical evolution of hot spots
with the aid of cocoon dynamics (Begelman \& Cioffi 1989: hereafter BC89; Kino \& Kawakatu 2005: hereafter KK05).
Specifically,  the evolution of the hot spot velocity 
($v_{\rm HS}$), the hot spot pressure ($P_{\rm HS}$) and the hot spot density 
($\rho_{\rm HS}$) are discussed.
These quantities are described 
in terms of the length from the center of the 
galaxy to the hot spot ($l_{\rm h}$).

Concerning $v_{\rm HS}$,  
radio observations of powerful FR II radio galaxies show us that
hot spots are always reside at the tip of the radio lobe 
(e.g., Myers \& Spangler 1985; Readhead et al. 1996b).
Thus, it is natural to impose the relation of 
\begin{eqnarray}
v_{\rm HS}=v_{\rm h},
\end{eqnarray}
where $v_{\rm h}$ is the advance speed of the cocoon head. 
%
%
The velocity $v_{\rm h}$ is significantly
affected by the two dimensional (2D) effect. 
However it can be reasonably handled by the 
phenomenological description as follows. 
Consider a pair of jets propagating in an ambient 
medium (see Fig. 1). At the hot spot, the 
flow of the shocked matter is spread out by the oblique shocks 
that then deflects (Lind et al. 1989), the vortex occurs via 
shocks (e.g., Smith et al. 1985) and/or the effect of jittering 
of the jet (e.g., Williams \& Gull 1985; Cox et al. 1991) which 
behaves like the ``dentist drill'' (Scheuer 1982). Thus, the
 effective ``working surface'' for the advancing jet is 
larger than the cross section area of the hot spot $A_{\rm j}$,
which was pointed out by BC89.
BC89 introduced the effective cross section area of the 
cocoon head $A_{\rm h}$ as that of the effective ``working surface''. 
\footnote{Before BC89, the head advance velocity is estimated by purely
 the 1D momentum balance (e.g., Begelman, Blandford, and Rees 1984). }
Thus, we can determine the reasonable value of 
 $v_{\rm h}$ by the expanding cocoon process.

As for $P_{\rm HS}$ and $\rho_{\rm HS}$,
we deal with them through one dimensional (1D) shock junctions. 
Since the hot spot is identified with the 
reverse shocked region of the jet, $P_{\rm HS}$ and $\rho_{\rm HS}$ 
can be obtained as a function of $v_{\rm h}$ by combining with eq. (1).

\begin{figure}
\includegraphics[width=8cm]{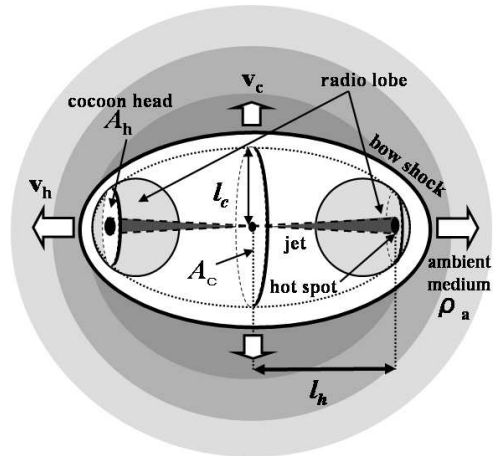}
\caption
{A schematic picture of the co-evolution of hot spots and cocoons. 
Most of the kinetic energy jet is 
injected via the termination shock of the jet
which is identified as the hot spot.
The sideways expansion speed of cocoon is $v_{\rm c}$.
The area of the radio lobe at the position of hot 
spots $A_{\rm h}$ is larger than that of hot spots. 
The head part of the cocoon advances with speed $v_{\rm h}$. }
\end{figure}

\subsection{Un-shocked Jet}

Here we set up the 
mass, momentum and energy flux of the 
jet with three assumptions.
Our main assumptions are as follows;
(i)
We assume that
the speed of the jet is relativistic on 
the large scale ($\sim 100\,{\rm kpc}$)  and the jet 
is consist of the cold medium.
Although the jet speed on large scales is still open issue,
several jets are suggested to be  
relativistic ones (e.g., Tavecchio et al. 2000; Celotti, 
Ghisellini \& Chiaberge 2001; Uchiyama et al. 2005),
(ii) The mass, energy and momentum of jets are conserved in time. 
Namely, we do not include the entrainment effect
of the ambient medium.
%
This is justified by the numerical simulations 
for highly relativistic jet flows (e.g., Scheck et al. 2002; 
Mizuta, Yamada \& Takabe 2004) and
(iii)
We ignore the dynamical effect of magnetic fields 
because of the kinetic flux dominance in FR II radio galaxies 
(e.g., Hardcastle \& Worrall 2000; Leahy \& Giani 2001; Isobe et al. 2002). 
Then, 
the mass ($J_{\rm 1D}$), energy ($L_{\rm 1D}$)  
and momentum ($Q_{\rm 1D}$) flux are given (Blandford \& Rees 1974);
\begin{equation}
J_{\rm 1D}=\Gamma_{\rm j}A_{\rm j}\rho_{\rm j}c,
\end{equation}
\begin{equation}
L_{\rm 1D}=\Gamma_{\rm j}^{2}A_{\rm j}\rho_{\rm j}c^{3},
\end{equation}
\begin{equation}
Q_{\rm 1D}=\Gamma_{\rm j}^{2}A_{\rm j}\rho_{\rm j}c^{2},
\end{equation}
where 
$\Gamma_{\rm j}$ and $\rho_{\rm j}$ 
are the Lorentz factor and the mass density of the jet, respectively. 
Note that the kinetic energy flux $L_{\rm 1D}$ denoted here
satisfies the relation of $L_{\rm j}=(A_{\rm h}/A_{\rm j})L_{\rm 1D}$, 
where $L_{\rm j}$ is the total kinetic power shown in KK05.

From these conditions, the following quantities are conserved for any $l_{\rm h}$;
\begin{equation}
\Gamma_{\rm j}={\rm const}, 
\end{equation}
\begin{equation}
\rho_{\rm j}A_{\rm j}\,c={\rm const}.
\end{equation}
Interestingly, the Lorentz factor $\Gamma_{\rm j}$ does not depend on $l_{\rm h}$. In other words, the speed of jet is relativistic even on the large scale.

\subsection{Shock junctions between the jet and  ambient medium} 
We briefly review the 1D shock jump 
conditions which governs the deceleration of  
the relativistic jet by the surrounding ambient medium
(Kino \& Takahara 2004 for details).
%
%
%
%
%
%
We can determine
$P_{a}$ and $\rho_{a}$ 
from X-ray observations
where $P_{\rm a}$ and $\rho_{a}$ are the pressure and the mass density of the ambient medium, respectively.
The assumption of a cold jet is written as $P_{\rm j}=0$. 
We regard $\Gamma_{\rm j}$ 
as a parameter. 
%

Since $v_{\rm HS}$ is estimated to be in the range 
0.01c to 0.1c both for FR II sources (Liu, Pooley \& Riley 1992; 
Scheuer 1995) and CSOs (Conway 2002 and references therein), 
the forward shocked (FS) region quantities are determined by the 
shock jump conditions in non-relativistic limit (Landau \& Lifshitz 1959).
By using the pressure balance along the contact
discontinuity between the hot spot and ambient medium,
we can obtain the expression of  $P_{\rm HS}$ as functions of two 
observable quantities $v_{\rm HS}$ and $\rho_{\rm a}$ such as  
\begin{equation}\label{eq:Phs}
P_{\rm HS}=\frac{4}{15}
\frac{[5-(1/{\cal M}^{2})]}{[1-(1/{\cal M}^{2})]^{2}}
\rho_{\rm a}v_{\rm HS}^{2},
\end{equation}
where 
${\cal M}=v_{\rm FS}
/\sqrt{(5P_{\rm a}/3\rho_{\rm a})}$ and $v_{\rm FS}$ are 
the Mach number and the velocity of the upstream of FS, respectively.
We adopted the adiabatic index of the downstream of FS as 5/3. 
%
In the reverse shocked (RS) region, 
we employ the relativistic shock jump conditions 
in the strong shock limit (Blandford \& McKee 1976).
Then, the equation of state and 
the mass continuity in the RS region 
can be written as
\begin{equation}
\rho_{\rm HS}=\frac{3P_{\rm HS}}{(\Gamma_{\rm j}-1)c^{2}},
\end{equation}
\begin{equation}
\rho_{\rm j}=\frac{3P_{\rm HS}}{(4\Gamma_{\rm j}+3)
(\Gamma_{\rm j}-1)c^{2}},
\end{equation}
where we set 
the adiabatic index in the RS region as $4/3$. 
Thus, $\rho_{\rm HS}$ and $\rho_{\rm j}$ also can be 
given by $\rho_{\rm a}$ and $v_{\rm HS}$.

\subsection{Dynamical evolution of the cocoon}

To determine the velocity of the cocoon head $v_{\rm h}$ with 
considering 2D sideways expansion, we prepare 
the solutions of cocoon dynamics based on KK05.
In KK05, by solving the equation 
of motion along the jet axis, 
perpendicular to the axis (i.e., sideways expansion), 
and energy injection into the cocoon, 
we obtained the 
$v_{\rm c}$, $v_{\rm h}$, $P_{\rm c}$, and $A_{\rm h}$ 
in terms of $l_{\rm h}$, where $v_{\rm c}$, and $P_{\rm c}$ 
are the velocity of cocoon sideways expansion and 
the pressure of cocoon, respectively. 
The declining mass density of the ambient medium is assumed to be $\rho_{\rm a}(d)=\rho_{\rm a0}(d/d_{0})^{-\alpha}$, where $d$, $d_{\rm 0}$ and $\rho_{\rm a0}$ are the radial distance from the center of the galaxy, 
the reference position and the mass density of the ambient medium at $d_{0}$, respectively.
In order to convert $t$-dependence of the results of 
KK05 to $l_{\rm h}$-dependence, we use the equation 
$l_{\rm h}=\int_{0}^{t} v_{\rm h}(t') d t'$ and $l_{\rm c}=\int_{0}^{t} v_{\rm c}(t') d t'$ 
where 
$l_{\rm c}$ is the radius of the cocoon body.
The obtained cocoon quantities in KK05 are as follows; 
\begin{equation}
v_{\rm c}=v_{{\rm c}0}\left(\frac{l_{\rm h}}{l_{{\rm h}0}}\right)^{\frac{0.5X-1}{X(-2+0.5\alpha)+3}}, 
\end{equation}
\begin{equation}
P_{\rm c}=P_{{\rm c}0}\left(\frac{l_{\rm h}}{l_{{\rm h}0}}\right)^{\frac{X(1-\alpha/2)-2}{X(-2+0.5\alpha)+3}}, 
\end{equation}
\begin{equation}
v_{\rm h}=v_{{\rm h}0}\left(\frac{l_{\rm h}}{l_{{\rm h}0}}\right)^{\frac{2-X(2-0.5\alpha)}{X(-2+0.5\alpha)+3}},  
\end{equation}
\begin{equation}
A_{\rm h}=A_{{\rm h}0}\left(\frac{l_{\rm h}}{l_{{\rm h}0}}\right)^{\frac{X(\alpha-2)(-2+0.5\alpha)+3\alpha-4}{X(-2+0.5\alpha)+3}}, 
\end{equation}
where $X$ is the power law index of the effective cross section area of the cocoon body $A_{\rm c}(t)\propto t^{X}$ (see Fig.1). 
Throughout this paper, we use the normalization
of $l_{{\rm h}0}=v_{{\rm h}0}t_{7}$, where $t_{7}=10^{7}{\rm yr}$ 
and $v_{{\rm h}0}$ is the velocity of the cocoon head at $l_{{\rm h}0}$ 
and the coefficients of each physical quantities 
are denoted with the subscript 0. 
In KK05, we selected $A_{\rm c}$ as an unknown parameter because we could not obtain the solution for $\alpha=2$. However, by compare with previous works, it is worth to show the power law index $\beta$ of the effective cross section area of the cocoon head $A_{\rm h}=A_{{\rm h}0}(l_{\rm h}/l_{{\rm h}0})^{\beta}$ as a function of $\alpha$ and $X$. 
From eq. (13), we obtain the following relation as 
\begin{equation}
\beta = \frac{X(\alpha-2)(-2+0.5\alpha)+3\alpha-4}{X(-2+0.5\alpha)+3}.
\end{equation}
From that, it is clear that the head growth can be expressed accurately 
in KK05 even for $\alpha=2$. 

Self-similar models are also useful tools to explore the evolution 
of the cocoon expanding (e.g., Falle 1991; Begelman 1996; Kaiser \& Alexander 1997). 
However, the problem has been pointed out on the 
assumption of the constant aspect ratio of cocoon employed 
in self-similar models. 
By the comparison of the young and grown-up sources,
they claim that the condition of the constant 
aspect ratio is not fulfilled incidentally
(e.g, Readhead et al. 1996a; De Young 1997; Komissarov \& Falle 1998; 
O'Dea 1998; Scheck et al. 2002; Carvalho \& O'Dea 2002a, b; 
Tinti \& De Zotti 2006).

%
%

\subsection{Dynamical evolution of hot spots}


Combining with 
(i) the mass, momentum, and kinetic energy of the jet ($\S 2.2$), 
(ii) the deceleration process of the jet by shock ($\S 2.3$),
 and 
(iii) the cocoon expansion ($\S 2.4$), 
we can finally obtain the dynamical evolution of hot spots. 
From eqs. (1), (5), (7), (8), (9) and (12), the quantities
$v_{\rm HS}$, 
$P_{\rm HS}$, 
$\rho_{\rm HS}$, and $\rho_{\rm j}$ as follows; 
\begin{equation}
v_{\rm HS}
=v_{{\rm HS}0}
\left(\frac{l_{\rm h}}{l_{{\rm h}0}}\right)^{\frac{2-X(2-0.5\alpha)}{X(-2+0.5\alpha)+3}},
\end{equation}
\begin{equation}
P_{\rm HS}=P_{{\rm HS}0}\left(\frac{l_{\rm h}}{l_{{\rm h}0}}\right)^{\frac{X(2-0.5\alpha)(\alpha-2)+4-3\alpha}{X(-2+0.5\alpha)+3}}, 
\end{equation}
\begin{equation}
\rho_{\rm HS}=\rho_{{\rm HS}0}\left(\frac{l_{\rm h}}{l_{{\rm h}0}}\right)^{\frac{X(2-0.5\alpha)(\alpha-2)+4-3\alpha}{X(-2+0.5\alpha)+3}}, 
\end{equation}
\begin{equation}
\rho_{\rm j}=\rho_{{\rm j}0}\left(\frac{l_{\rm h}}{l_{{\rm h}0}}\right)^{\frac{X(2-0.5\alpha)(\alpha-2)+4-3\alpha}{X(-2+0.5\alpha)+3}}. 
\end{equation}
Note that $P_{{\rm HS}0}$, $\rho_{{\rm HS}0}$ and 
$\rho_{{\rm j}0}$ can be expressed by only observable quantities $\rho_{{\rm a}0}$ and $v_{{\rm HS}0}$ if we assume $\Gamma_{\rm j}$ (see eqs. (7), (8) and (9)).
Thus, it is possible to know not only $l_{\rm h}$ dependence but also the absolute quantities of hot spots and jets.
%
The aspect ratio of the cocoon 
${\cal R}\equiv l_{\rm c}/l_{\rm h}$ is the intriguing quantity for 
studying the dynamical evolution of hot spots.
The $l_{\rm h}$-dependence of the aspect ratio of cocoon is 
then given by 
\begin{equation}
{\cal R}={\cal R}_{0}\left(\frac{l_{\rm h}}{l_{{\rm h}0}}\right)^{\frac{X(2.5-0.5\alpha)-3}{X(-2+0.5\alpha)+3}}, 
\end{equation}
where ${\cal R}_{0}=(v_{{\rm c}0}/v_{{\rm HS}0})[(X(-2+0.5\alpha)+3)/(0.5X)]$. 
%
%
%
As a consistency check of our assumption of constant 
$A_{\rm h}/A_{\rm j}$, we can easily find that hot spot radius 
$r_{\rm HS}$ ($\propto A_{\rm j}^{1/2}$) shows the same 
$l_{\rm h}$-dependence as $A_{\rm h}^{1/2}$ from eqs. (6) and (18).
From above results, we obtain the slope of all physical quantities 
as functions of two key physical quantities, namely $\alpha$ 
(the slope index of the ambient matter 
density) and X (the growth rate 
of cross section of cocoon body).
In the case of constant ${\cal R}$, our results agree with self-similar models of cocoon expansions (e.g., Falle 1991; Begelman 1996; Kaiser \& Alexander 1997). However, we stress that these self-similar models assume that 
$P_{\rm HS}/P_{\rm c}$ and ${\cal R}$ are both constant 
in $l_{\rm h}$, whilst we do not impose these assumptions 
and also predict the dynamical evolution of $\rho_{\rm HS}$ and $\rho_{\rm j}$.  

The relation between $P_{\rm HS}$
and $P_{\rm c}$ is also the interesting topic.
From eq. (\ref{eq:Phs}), the hot spot pressure is 
written by $P_{\rm HS}=4\rho_{\rm a}(l_{\rm h})v_{\rm HS}^{2}/3$ 
for ${\cal M} \gg 1$, while the 
over-pressured cocoon requires
$P_{\rm c}=\rho_{\rm a}(l_{\rm c})v_{\rm c}^{2}$.
Thus, the ratio of $P_{\rm HS}$ to $P_{\rm c}$ is 
\begin{equation}\label{eq:ratioP}
\frac{P_{\rm c}}{P_{\rm HS}}=\frac{3}{4}
\left[\frac{0.5X}{X(-2+0.5\alpha)+3}\right]^{2}{\cal R}_{0}^{\alpha}
{\cal R}^{2-\alpha}.
\end{equation}
This implies that
$P_{\rm c}/P_{\rm HS}$ is controlled by 
${\cal R}$ and $\alpha$. 
Since  ${\cal R}<1$ and ${\cal R}_{0}<1$ are satisfied
by definitions, $P_{\rm c}/P_{\rm HS}$ should
less than unity for $0 < \alpha < 2$. 
In the case of ${\cal R}={\cal R}_{0}={\rm const}$ 
or $\alpha=2$, it reduces to the interesting relation of 
\begin{eqnarray}
P_{\rm c}/P_{\rm HS}=\frac{3}{4}{\cal R}^{2}_{0}. \nonumber
\end{eqnarray}
This shows that $P_{\rm c}/P_{\rm HS}$ is 
determined only by ${\cal R}^{2}_{0}$.
We stress that our model predict that 
$P_{\rm c}$ is smaller than
$P_{\rm HS}$ as long as ${\cal R}_{0} < 1$.
Additionally, rewriting of 
the explicit form of the $P_{\rm c}$ 
in terms of the quantities of the jet may be also 
stimulating, which is given by
$P_{\rm c}=
{\cal R}^{2}_{0}
\Gamma_{\rm j}^{2}
\rho_{\rm j}c^{2}$
for ${\cal R}_{\rm 0}={\rm const}$.
From this, one can find that 
the larger $\Gamma_{\rm j}$ leads to the larger 
$P_{\rm c}$ which is actually 
seen in relativistic hydrodynamic 
simulations (see Fig. 5 in Mart\'{i} et al. 1997). 
To comprehend  the energy injection process
into the cocoon via the hot spot
with the duration of  $t_{\rm inj}$, 
we rewrite the eq. (\ref{eq:ratioP}) as
\begin{eqnarray}
P_{{\rm c}0}v_{{\rm c}0} S_{{\rm c}0}\approx
P_{{\rm HS}0}v_{{\rm esc}0} S_{{\rm HS}0}\equiv L_{\rm j}t_{\rm inj}, 
\end{eqnarray}
where 
$S_{{\rm HS}0}\equiv 4\pi r_{{\rm HS}0}^{2}$, and
$S_{{\rm c}0}\equiv 2\pi l_{{\rm c}0}l_{{\rm h}0}$,
$v_{{\rm esc}0}\equiv c/(2(0.5X)^{2})\sim (0.5-0.7)c$.
This describes the  
{\it continuous} energy injection
of AGN jets (i.e., $t_{\rm inj}=t_{\rm age}$).
On the contrary,
Blandford and Rees (1974) used the relation of
$P_{{\rm c}0}(v_{{\rm c}0} S_{{\rm c}0})^{\hat \gamma_{c}}
\approx
P_{{\rm HS}0}(v_{{\rm esc}0} S_{{\rm HS}0})^{\hat \gamma_{HS}}$
where ${\hat \gamma}$ is the adiabatic index in each region.
We claim that this relation is appropriate  
for the instantaneous (i.e., $t_{\rm inj}\ll t_{\rm age}$)
injection  seen in supernovae (SNe) or gamma-ray bursts (GRBs).

\section{Comparison with previous works}

\begin{table}
\begin{center}
{Table 1. Comparison with 2D hydrodynamic simulations and self-similar models}
{
\begin{tabular}{ccccccc}
\hline \hline
----- &  $v_{\rm HS}$ & $A_{\rm h}$ &$P_{\rm c}$ & $P_{\rm HS}$ & $\rho_{\rm j}$& ${\cal R}$ \\
\hline
``1D'' Phase$^{a}$ & & & &  & & \\
S02 & $l_{\rm h}^{-0.11}$ & const & $l_{\rm h}^{-0.95}$ & const & const & $l_{\rm h}^{-0.45}$ \\ 
\\
BC89 & const & const & $\l_{\rm h}^{-1}$ & const &-----& $l_{\rm h}^{-0.5}$ \\
\\
This work & const & const & $l_{\rm h}^{-1}$ & const & const & $l_{\rm h}^{-0.5}$ \\
\hline
``2D'' Phase$^{b}$ & & & &  & \\
S02 & $l_{\rm h}^{-0.55}$ & $l_{\rm h}^{0.90}$ & $l_{\rm h}^{-1.30}$ & $l_{\rm h}^{-1.1}$ &$l_{\rm h}^{-1.0}$ & $l_{\rm h}^{-0.09}$ \\ 
\\
B96 & $l_{\rm h}^{-2/3}$ & $l_{\rm h}^{4/3}$ & $l_{\rm h}^{-4/3}$ & $l_{\rm h}^{-4/3}$ &----- & const \\
\\
This work & $l_{\rm h}^{-0.56}$  & $l_{\rm h}^{1.1}$ & $l_{\rm h}^{-1.30}$ & $l_{\rm h}^{-1.1}$ &$l_{\rm h}^{-1.1}$ & $l_{\rm h}^{-0.08}$ \\\hline
\end{tabular}
}
\noindent
\end{center}
{NOTE.--
$^{a}$The 1D phase  corresponds to our model with 
$\beta=0$ and $\alpha=0$. 
$^{b}$The 2D phase (b) corresponds to our model 
with $\beta=1.1$ and $\alpha=0$.}
\end{table}

\subsection{Comparison with numerical simulations}
Scheck et al. (2002; hereafter S02) examined
the long term evolution of the powerful jet propagating 
into a uniform ambient medium ($\alpha=0$)
with ``2D'' relativistic hydrodynamic simulations.
S02 showed that the evolution of the jet proceeds in two different 
phases appear (they are shown in Table 4 and Fig. 6 in S02).
{\it ``1D'' phase}: In the initial phase ($t < 1.2\times 10^{5}{\rm yr}$), 
the jet shows ballistic propagation with 
$A_{\rm h}={\rm const}$ and $v_{\rm HS}={\rm const}$.
{\it ``2D'' phase}: 
This phase starts when the first large vortices are produced 
near the tip of the jet. 
Then, the cross section area of the cocoon 
head begins to increase. 
The hot spot then starts decelerating, but 
the jet speed remains the same relativistic one 
during whole simulations. 
Below we compare of the present work with the hydrodynamic 
simulation of S02 in Table 1.

In the ``1D'' phase, the results of S02 can be well
described by our model with $\beta=0$ and $\alpha=0$. 
Note that  this ``1D'' phase corresponds to the evolutionary 
model with constant $A_{\rm h}$ (BC89).
For $v_{\rm HS}$, the power law index is slightly ($\sim 10\%$,) 
different from our model (also BC89) and the results of S02. 
In this case, $P_{\rm c}\propto l_{\rm h}^{-1}$ and $P_{\rm HS}={\rm const}$ 
are predicted by this work and BC89, which coincides with the numerical 
results of S02 (see Fig. 6 (c) for $P_{\rm c}$ and $P_{\rm HS}$ in S02).  
In addition, our model can reproduce the constant $\rho_{\rm j}$ 
(see Fig. 5(a) in S02). 
For comparisons, let us briefly comment on the self-similar
model of expanding cocoons in which the growth of the 
cocoon head is included (e.g., Begelman 1996: hereafter B96). 
As already pointed out (e.g., Carvalho \& O'Dea 2002), the 
self similar mofel of B96 cannot represent   the behavor of
the ``1D'' phase. 
The behavior of $P_{\rm c}/P_{\rm HS}$ is also the intriguing issue. 
The decrease of $P_{\rm c}/P_{\rm HS}$  with time 
is reported in Fig. 6 of S02. 
Using our model, this behavior is clearly
explained by the decrease of the 
cocoon aspect ratio (see eq. (20)).

The ``2D'' phase of S02 is well described by
our model with $\beta=1.1$ and $\alpha =0$.  
We adopt $\beta =1.1$ to reproduce the $P_{\rm c}$ evolution 
in Fig. 6 (c) of S02 because the other quantities shows much larger 
fluctuations in Fig.6 of S02.
The present model predicts the evolution 
of the hot spot pressure and mass density of the jet 
as $P_{\rm HS}\propto l_{\rm h}^{-1.1}$, $v_{\rm HS}\propto 
l_{\rm h}^{-0.56}$ and $\rho_{\rm j}\propto l_{\rm h}^{-1.1}$. 
These coincides with the average value of $P_{\rm HS}$, $v_{\rm HS}$, 
and $\rho_{\rm j}$ (see Fig.5 (a) and Fig 6 in S02). 
In the ``2D'' phase, the cross section of cocoon head grows as 
$A_{\rm h}\propto l_{\rm h}^{1.1}$ unlike the ``1D'' phase 
($A_{\rm h}={\rm const}$). Thus, the velocity 
of hot spot decreases with $l_{\rm h}$. 
Actually, the growth of the cross section area of the cocoon 
head can be seen in their simulations (Fig. 3 in S02). 
In this phase, B96 also explains these results of S02. 
Moreover, the cocoon pressure is 
proportional to $P_{\rm HS}$ in this phase of S02. 
From eq. (20),
it can be understood with a constant  ${\cal R}$. 
From above detailed comparison with ``2D'' relativistic 
hydrodynamic simulations, we found that the
model represented in this paper can 
describe the flow and cocoon behavors seen in the ``1D'' 
and ``2D'' phases  very well. 
It should be stressed that 
our analytic model is more useful than  numerical simulations 
when investigating a longer-term evolution of jets. 
The length of jets performed by  numerical  
simulations of jets achieves at most the length
order of $100$ times of a jet beam size,
while the spacial sizes of actual jets in AGNs are spread 
in six order of magnitude (i.e., from parsec to mega-parsec scale).

\subsection{Comparison with observations}
Based on a number of recent reports of 
indicating that the constant speed of 
hot spot advance (0.01 $< v_{\rm HS}/c < $0.1) 
(e.g., Readhead et al. 1996b;  Carilli et al. 1991; 
Conway 2002), we here examine 
the case of $v_{\rm HS}={\rm const}$. 
Observationally, $P_{\rm HS}$ and $\rho_{\rm HS}$ are inferred 
by using the minimum energy assumption and the neglecting the effect of 
thermal components (Readhead et al. 1996a; Jeyakumar \& Saikia 2000; 
Perucho \&  Mart\'{i} 2002).  
From eq. (15), the relation of $2-X(2-0.5\alpha)=0$ is required 
for the constant hot spot velocity.
By eliminating the parameter $X$, 
our model reduce to the following forms;
$v_{\rm c}\propto l_{\rm h}^{-(\alpha-2)/(\alpha-4)}$, 
$P_{\rm c}\propto l_{\rm h}^{4/(\alpha-4)}$, 
$P_{\rm HS}\propto l_{\rm h}^{-\alpha}$, 
$\rho_{\rm HS}\propto l_{\rm h}^{-\alpha}$, 
$ r_{\rm HS}\propto l_{\rm h}^{\alpha/2}$, 
$ \rho_{\rm j}\propto l_{\rm h}^{-\alpha}$, and 
$ {\cal R}\propto l_{\rm h}^{-(\alpha-2)/(\alpha-4)}$. 
Here we used mean density profiles obtained by a large  
number of sample clusters of galaxies, which is 
$\rho_{\rm a}(d)\propto d^{-(1.5 \textrm{--} 2)}$ 
(e.g., Mulchaey \& Zabludoff 1998).

We show the comparison with observational data for CSOs 
and FR II sources in Table 2.
This indicates that 
our model well reproduce observational trends 
within the error bars. 
These agreements are likely to support 
``youth radio source scenario" basically. 
At the same time, 
the large dispersion of the observational data
could tell us other  possibilities of
evolutionary tracks of radio loud AGNs  
usually discussed. 
To explore it, it must be valuable to inquire into 
the origin of their large dispersion. 
Furthermore, we emphasize that 
the deviation from the self similar evolution 
are frequently indicated by several authors 
(e.g., De Young 1997, Gilbert et al. 2004).

It is worth to show the reliability 
of the relation of the opening angle of hot spots derived by eq. (21), namely 
$\theta_{\rm HS}=(v_{\rm HS}/c)^{1/2}{\cal R}^{2}_{0}$. 
For this aim, we adopt this equation to an archetypal 
radio galaxy Cygnus A. 
Using the values of
$v_{\rm HS}\approx 0.01c$ (Carilli et al. 1991)
and 
${\cal R}=0.6$ (Wilson, Young, \& Shopbell 2000) 
our model predicts 
$\theta_{\rm HS}\simeq 0.036$, 
while the direct estimate of $\theta_{\rm j}$ 
with $r_{\rm HS}=2{\rm\, kpc}$ and $z=60{\rm\, kpc}$
indicate $\theta_{\rm HS}\simeq0.033$.
Thus we can verify the reliability of the relation for 
the opening angle of hot spots and then we propose a new way 
of the estimation of $v_{\rm HS}$ from obsevable two quantities ${\cal R}$ and $\theta_{\rm HS}$.
It would be worth to compare with evaluations from the kinematic studies.

\begin{table}
\begin{center}
{Table 2. Comparisons with observations}
{
\begin{tabular}{ccccc}
\hline \hline
----- &  $v_{\rm HS}$ & $P_{\rm HS}$ & $r_{\rm HS}$ & $\rho_{\rm HS}$ \\
\hline
Observations$^{a}$ & const & $l_{\rm h}^{-(1.3 \textrm{--}1.7)}$  & $l_{\rm h}^{ 0.7 \textrm{--}1.3}$ & $l_{\rm h}^{-(1.9 \textrm{--}2.9)}$\\ \\
This work$^{b}$ & const & $l_{\rm h}^{-(1.5 \textrm{--}2.0)}$ & $l_{\rm h}^{0.75\textrm{--}1}$ & $l_{\rm h}^{-(1.5 \textrm{--}2.0)}$\\
\hline
\end{tabular}
}
\noindent
\end{center}
{NOTE.--
$^{a}$
The results are adopted from 
Readhead et al. (1996a), 
Jeyakumar \& Saikia (2000)
and Perucho \&  Mart\'{i} (2002).
$^{b}$We set the slope index of the 
ambient density $\alpha=1.5\textrm{--}2$. }
\end{table}

%
%


\section{Conclusions} 
In the present work, we model a dynamical evolution of hot spots 
in radio loud AGNs. 
In this model, the unshocked flow
satisties
the conservations of the mass, momentum, and kinetic energy. 
We take account of  
the deceleration process of the jet by shocks, and 
the cocoon expansion which is identified as the 
by-product of the exhausted flow.  
The model describes 
the evolution of physical quantities ($v_{\rm HS}$, $P_{\rm HS}$, 
and $\rho_{\rm HS}$) in the hot spot in terms of $l_{\rm h}$. 
%
%
%
%
Below we summarize the main results based on this model.

\begin{enumerate}

\item 
We find that the ratio of $P_{\rm c}/P_{\rm HS}$
is controlled by the aspect ratio 
of the cocoon ${\cal R}$ and slope index of the 
ambient medium $\alpha$. 
If ${\cal R}$ remains to be constant
during the jet propagation, the value $P_{\rm c}/P_{\rm HS}$
is proportional to ${\cal R}^{2}$ 
with the coefficient of order unity. 
This naturally explain the basic concept 
of the elongated cocoon with larger $P_{\rm HS}$ than $P_{\rm c}$.
Concerning $P_{\rm c}$, 
it is proportional to the bulk kinetic power of the jet 
in given $\rho_{\rm a}$. 
This is originated from our treatment of adiabatic injection
of the dissipation energy of the jet into the cocoon.
In addition, we suggest a new method to evaluate the velocity of 
hot spots from the aspect ratio of cocoon and the opening angle of 
hot spots.

\item Our analytic model can well explain the results of 
2D co-evolution of jets and cocoons obtained by relativistic 
hydrodynamic simulations. 
This clearly guarantees the reliability of our model
during the over-pressure cocoon phase.
Since the dynamical length of jets obtained by 
numerical simulations is 
about a few $100$ times of the jet beam size, 
our analytic model must be an useful tool 
to explore a longer-term dynamical evolution of jets than this scale. 
%

%

\item Our model prediction
reasonably coincides with the recent observational trends 
of hot spots seen in CSO and FR II sources. 
Furthermore, we predict 
${\cal R}\propto l_{\rm h}^{-(0.2 \textrm{--} 0)}$ and 
$A_{\rm h}\propto l_{\rm h}^{1.5 \textrm{--} 2}$ 
although little is done about systematic studies on
these quantities.
%


Lastly
we should keep in mind that the present model
do not take account of the details of 
(i) the absolute value of the mass density of 
the ambient medium, 
and 
(ii) radiative cooling effect which may be important for
younger radio galaxies. 
In order to investigate whole story of evolutionary track
of the radio loud AGNs, the study of 
above two points will be inevitably required.
We plan investigate both of them in the near future.

\end{enumerate} 


\section*{Acknowledgments}
We thank A. Celotti, H. Ito and F. Takahara 
for useful discussions and comments.
We acknowledge the Italian MIUR and INAF financial support. 
We also thank an anonymous referee for helpful comments 
to improve this papaer.

\label{lastpage}


\begin{thebibliography}{99}
\bibitem{} Begelman M. C., 1996, in Cygnus A---Study of a Radio Galaxy, ed. C. L.Carilli \& D. E. Harris (Cambridge: Cambridge Univ. Press), 209
\bibitem{} Begelman M. C., Cioffi, D. F., 1989, ApJ, 345, L21(BC89)
\bibitem{} Begelman M.~C., Blandford R.~D., Rees M.~J., 1984, RvMP, 56, 255
\bibitem{} Blandford R. D., McKee C. F., 1976, Physics of Fluids, 19, 1130
\bibitem{} Blandford R. D., Rees M. J., 1974, MNRAS, 169,395
\bibitem{} Bridle A. H., Perley R. A., 1984, ARA\&A, 22, 319
\bibitem{} Carvalho J.~C., O'Dea C.~P., 2002a, ApJS, 
141, 337
\bibitem{} Carvalho J.~C., O'Dea C.~P., 2002b, ApJS, 
141, 371
\bibitem{} Carvalho J. C., 1985, MNRAS, 215, 463
\bibitem{} Carilli C. L., et al., 1991, ApJ, 383, 554
\bibitem{} Celotti A., Ghisellini G., Chiaberge, M., 2001, MNRAS, 321, L1
\bibitem{} Clarke D. A., Norman M. L., Burns J. O., 1986, ApJ, 311, L63
\bibitem{} Clarke D. A., Harris D. E., Carilli C. L., 1997, MNRAS, 284, 981
\bibitem{} Conway J. E., 2002, NewAR, 46, 263
\bibitem{} Cox C. I., Gull S. F., Scheuer P. A. G., 1991, MNRAS, 252, 558
\bibitem{} De Young D. S., 1997, ApJ, 490, L55
\bibitem{} Falle S. A. E. G., 1991, MNRAS, 250, 581
\bibitem{} Fanti C., et al., 1995, A\&A, 302, 317
\bibitem{} Gilbert G.~M., Riley J.~M., Hardcastle 
M.~J., Croston J.~H., Pooley G.~G., Alexander P., 2004, MNRAS, 351, 845
\bibitem{} Hardcastle M.~J., Worrall D.~M., 2000, MNRAS, 319, 562
\bibitem{} Isobe N., et al., 2002, ApJ, 580, L111
\bibitem{} Jeyakumar S., Saikia D.~J., 2000, MNRAS, 311, 397
\bibitem{} Kaiser C. R., Alexander P., 1997, MNRAS, 286, 215
\bibitem{} Kino M., Takahara F., 2004, MNRAS, 349, 336
\bibitem{} Kino M., Kawakatu N., 2005, MNRAS, 364, 659 (KK05)
\bibitem{} Komissarov S.~S., Falle S.~A.~E.~G., 1998, MNRAS, 297, 1087
\bibitem{} Landau L. D., Lifshitz E. M., 1959, Fluid Mechanics, Pergamon Press, Oxford
\bibitem{} Leahy J. P., Gizani N. A. B., 2001, ApJ, 555, 709
\bibitem{} Lind K. R., et al., 1989, ApJ, 344, 89
\bibitem{} Liu R., Pooley G., Riley, J. M., 1992, MNRAS, 257, 545
\bibitem{} Mart\'{i} M. A., et al., 1997, ApJ, 479, 151
\bibitem{} Mizuta A., Yamada S., Takabe H., 2004,ApJ, 606, 804
\bibitem{} Mulchaey J. S., Zabludoff A.I., 1998, ApJ, 496, 73
\bibitem{} Murgia M., Fanti C., Fanti R., Gregorini L., Klein U., Mack K.-H., Vigotti M., 1999, A\&A, 345, 769 
\bibitem{} Myers S.T., Spangler S.R., 1985,ApJ, 291, 52
\bibitem{} Norman M. L., Smarr L., Winkler K-H. A., Smith M. D., 1982, A\&A, 113, 285
\bibitem{} O'Dea C.~P., 1998, PASP, 110, 493
\bibitem{} O'Dea C. P., Baum S. A., 1997, AJ, 113, 148
\bibitem{} Owsianik I., Conway J. E., Polatidis A.G., 1999, NewAR, 43, 669
\bibitem{} Perucho M., Mart\'{i} J. M., 2002, ApJ, 568, 639
\bibitem{} Phillips R. B., Mutel R. L., 1980, ApJ, 236, 89
\bibitem{} -----.1982, A\&A, 106, 21
\bibitem{} Readhead A. C. S., Cohen M. H., Blandford R. D., 1978, Nature, 272, 131
\bibitem{} Readhead A. C. S., et al., 1996a, ApJ, 460, 634
\bibitem{} Readhead A. C. S., et al., 1996b, ApJ, 460, 612
\bibitem{} Ryle M. S., Longair M.~S., 1967, MNRAS, 136, 123
\bibitem{} Scheck L., et al., 2002, MNRAS, 331, 615, 2002 (S02)
\bibitem{} Scheuer P. A. G., 1995, MNRAS, 277, 331
\bibitem{} Scheuer P. A. G., 1982, in Extragalactic Radio Sources, ed. D.S. Heeschen and 
C. M. Wade, IAU Symp. 97, Reidel Publishing Co., 163
\bibitem{} Shklovsky I. S. 1965, Nature, 206, 176
\bibitem{} Smith, M. D., et al. 1985, MNRAS, 214, 67
\bibitem{} Tavecchio F., et al. 2000, ApJ, 544, L23
\bibitem{} Taylor G.~B., Marr J.~M., Pearson T.~J., Readhead A.~C.~S., 2000, ApJ, 541, 112
\bibitem{} Tinti S., de Zotti G., 2006, A\&A, 445, 889
\bibitem{} Turland B. D., 1975, MNRAS, 172, 181
\bibitem{} Uchiyama Y., et al., 2005, ApJ, 63, L113 
\bibitem{} Wilkinson P. N., et al., 1994, ApJ, 432, L87
\bibitem{} Williams A. G., Gull F. G., 1985, Nature, 313, 34
\bibitem{} Wilson M.~J., Scheuer P.~A.~G., 1983, 
MNRAS, 205, 449
\bibitem{} Wilson A. S., Young A. J., \& Shopbell P. L., 2000, ApJ, 544, L27
\bibitem{} van Breugel W. J. M., Miley, G. K., Heckman, T. A. 1984, AJ, 89, 5
\end{thebibliography}
\end{document}